\documentclass[twocolumn,pre,aps,showpacs,preprintnumbers,amsmath,amssymb,superscriptaddress,floatfix]{revtex4}
\usepackage{epstopdf}
\usepackage{epsf} 
\usepackage{dcolumn}
\usepackage{bm}
\usepackage{latexsym}
\newcommand{\kav}{\langle k \rangle}
\newcommand{\xsize}{\epsfxsize=9 cm}
\begin{document}

\title{Accelerating consensus on co-evolving networks: the effect of committed individuals}
\author{P. Singh}
\affiliation{Dept. of Physics, Rensselaer Polytechnic Institute, 110 8th Street, Troy, NY 12180, USA}
\affiliation{Social and Cognitive Networks Academic Research Center, Rensselaer Polytechnic Institute,\\
110 8$^{th}$ Street, Troy, NY 12180--3590, USA}

\author{S. Sreenivasan\footnote{Corresponding author: sreens@rpi.edu}}
\affiliation{Dept. of Physics, Rensselaer Polytechnic Institute, 110 8th Street, Troy, NY 12180, USA}
\affiliation{Social and Cognitive Networks Academic Research Center, Rensselaer Polytechnic Institute,\\
110 8$^{th}$ Street, Troy, NY 12180--3590, USA}
\affiliation{Dept. of Computer Science, Rensselaer Polytechnic Institute, 110 8th Street, Troy, NY 12180, USA}

\author{B.K. Szymanski}
\affiliation{Social and Cognitive Networks Academic Research Center, Rensselaer Polytechnic Institute,\\
110 8$^{th}$ Street, Troy, NY 12180--3590, USA}
\affiliation{Dept. of Computer Science, Rensselaer Polytechnic Institute, 110 8th Street, Troy, NY 12180, USA}

\author{G. Korniss}
\affiliation{Dept. of Physics, Rensselaer Polytechnic Institute, 110 8th Street, Troy, NY 12180, USA}
\affiliation{Social and Cognitive Networks Academic Research Center, Rensselaer Polytechnic Institute,\\
110 8$^{th}$ Street, Troy, NY 12180--3590, USA}

\begin{abstract}
Social networks are not static but rather constantly evolve in time. One of the elements thought to drive the evolution of social network structure is homophily - the need for individuals to connect with others who are similar to them. In this paper, we study how the spread of a new opinion, idea, or behavior on such a homophily-driven social network is affected by the changing network structure. In particular, using simulations, we study a variant of the Axelrod model on a network with a homophilic rewiring rule imposed.  First, we find that the presence of homophilic rewiring within the network, in general, impedes the reaching of consensus in opinion, as the time to reach consensus  diverges exponentially with network size $N$. We then investigate whether the introduction of committed individuals who are rigid in their opinion on a particular issue, can speed up the convergence to consensus on that issue. We demonstrate that as committed agents are added, beyond a critical value of the committed fraction, the consensus time growth becomes logarithmic in network size  $N$. Furthermore, we show that slight changes in the interaction rule can produce strikingly different results in the scaling behavior of $T_c$. However, the benefit gained by introducing committed agents is qualitatively preserved across all the interaction rules we consider.


\end{abstract}

\pacs{87.23.Ge, 
89.75.Fb 
}
\maketitle

\section{Introduction}

A dynamical process occurring on a network can be strongly influenced by the evolution of the network's structure itself. Furthermore, if the dynamical process on the network directly affects the network's structural evolution, a complex feedback process arises. In the context of the spread of opinions, behaviors, or ideas on a social network, such an interplay between individual states and the network's structural evolution is expected on the basis of the theory of homophily.
Homophily, introduced by Lazarsfeld and Merton \cite{Lazarsfeld1954,McPherson2001}, describes the tendency of individuals to form social connections with those who are similar to them. Complementarily, the persistence of ties is also thought to depend strongly on the similarity of the individuals they connect \cite{Mergel2006,Burt2000}. If the traits of individuals are unchanging, then we expect that the structure of the network will stabilize when each link connects a pair of individuals who are sufficiently similar. However, if individuals influence one another to adopt new behaviors, opinions, or ideas, and thereby affect each other's attributes, then the mutual similarities between pairs of individuals can be thought of as continuously evolving entities. Thus, one can envision the structure of a social network as being in a constant state of flux: links between dissimilar individuals decay with time while new ties between similar individuals form at some rate. This continuous death and birth of links is presumably balanced in such a way that on average, at any given time, the mean number of connections ascribed to any individual is roughly constant, or at least, bounded from above \cite{Dunbar1992}. 

There have been few empirical studies which track the simultaneous {\it co-evolution} of network structure and individual behaviors.  Notably, Lazer et al. have recently studied \cite{Lazer2010} how political views of students in a public policy program, and the network structure of their interactions evolved over a two-semester observation period. The main finding of this study was that in the process of making connections to other individuals, homophilic selection on the basis of political views was weak, while race and religion-based selection was comparatively strong. Furthermore, the study found that an individual's political views at the end of the observation period were significantly correlated with the mean affiliation of his/her neighborhood at the beginning of the observation period (controlling for the individual's own initial views), an indication, possibly, of social influence.

In contrast, models of networks where social opinions and network structure co-evolve have been studied extensively in previous literature \cite{castellanoreview}.  Benczik et al. \cite{ BenczikEPL2008, BenczikPRE2009} studied a two-parameter voter model that could be tuned to study all cases between two extremes where nodes preferentially interacted with other nodes holding the same opinion, or those holding the opposite opinion. They demonstrated that three outcomes were possible depending on the parameter values - a consensus state, a disordered state, or a frozen, polarized state. Holme et al. \cite{Holme2006} studied a single-parameter model of a co-evolving network and demonstrated the existence of a non-equilibrium phase transition between a steady state with diverse co-existing opinions and a consensus state. Nardini et al. \cite{Nardini2008} studied a variant of the voter model where an individual, with a certain probability, either severs a tie with a neighbor whose state differs from its own and forms a new tie with another node, or otherwise adopts the neighbor's state. They showed how small changes to the interaction rules, such as the order of choosing the interacting individuals, as well as the introduction of an intermediate state in the voter model can dramatically affect the probability of consensus as well as consensus times. Finally, Vazquez et al. \cite{Vazquez2007} studied a model where nodes are assigned attributes and undergo influence as per the Axelrod model \cite{Axelrod1997}, while existing links are rewired with a probability proportional to the dissimilarity between the nodes they connect. They demonstrated that there arise three phases characterized by differences in the steady state network structure, as the number of possible traits per attribute is varied. A model similar to the one presented in \cite{Vazquez2007} was studied in  \cite{Centola2007}, where the authors showed how cultural diversity can be stably maintained despite the presence of cultural drift.

The network model we consider in this paper is similar to the latter
two studies \cite{Vazquez2007,Centola2007} in its use of an
Axelrod-type measure of the similarity between individuals which in turn
dictates how social influence and link rewiring occur.  However, the
central motivation of our work is to understand how a fast consensus
to a particular attribute can be ensured on such evolving networks.
In particular, we investigate how the introduction of {\it
committed agents} - individuals who are selectively immune to
influence on a given issue, and who hold the same opinion on that
issue - affects the evolution of opinions on the network as well as
network structure itself. The effect of committed individuals, all
holding the same opinion, has been previously studied on
structurally static networks \cite{Galam2007, Galam2008,
Zhang2011,Xie2011}.

The key finding in these studies was the existence of a critical
committed fraction below which the deterministic evolution equations
admit a mixed steady state in addition to the always present
consensus steady state, with a saddle point separating the two. As a
consequence, on finite networks \cite{Zhang2011,Xie2011}, the time
to attain consensus scales exponentially with system size when the committed fraction is below the
critical value. This is consistent with the known scaling of 
transition times between deterministically stable states in stochastic bistable
systems \cite{Hanggi1990}. In stark contrast, above the critical value of the committed fraction, the consensus process
is essentially deterministic with consensus times logarithmically dependent on network size, and where
the only steady-state solution to the deterministic equations is
the consensus state. Further analysis has also revealed \cite{Xie2011, Jeff2012} that this ``critical" point or threshold is a spinodal point associated with an underlying first-order transition \cite{Landau} in the phase diagram of the model.

In this article, we investigate whether a similar fast convergence to
consensus can be engineered through committed agents when the network structure is evolving in response to the spread of
opinions. More specifically, we aim to understand whether the
scaling behavior of consensus time with network size undergoes a
significant change as the committed fraction within a co-evolving network is monotonically
varied. The presence of committed
agents holding distinct opinions in the network obviously prevents
the system from attaining consensus, a situation previously studied
in \cite{ Mobilia2007,Biswas2009} - in the present work, we do not
consider this case.

\section{The Model}
In our model, individuals are represented by nodes on a network and every node is assigned a set of $F$ independent attributes that constitute the node's {\it state}. Each attribute can take one of $q$ distinct traits, represented by integers in $[0,q-1]$. Thus the node's state is represented by an $F$ component vector. Initially, each attribute of each node is assigned one of the $q$ values randomly.
The structure of the network connecting the individuals is initialized to an Erd\H{o}s-R\'enyi (ER) random network with a given average degree $\kav$. Next, we define the rules governing the evolution of individual states as well as the structure of the network connecting them. At each time step:
a node $i$ is selected at random, and one of its neighbors $j$ is selected at random. We then compute the {\it similarity} between nodes $i$ and $j$, where similarity is defined as the number of attributes for which $i$ and $j$ possess the same trait. Then,
for a randomly chosen attribute among the set of attributes on 
\begin{enumerate}
\item If the similarity is found to be equal to or above a given threshold $\phi$, node $j$ adopts the trait possessed by node $i$ for a randomly chosen attribute from among those for which they currently do not share the same trait. We refer to this as the {\it influence step}.
\item Otherwise, the link between $i$ and $j$ is severed and node $i$ randomly selects a node $k$ in the network from among those to which it is currently not connected, and forms a link to it. We refer to this as the {\it rewiring step}.
\end{enumerate}
Our model clearly is a variant of the Axelrod model with the following distinctions:  in the Axelrod model, the influence step occurs with probability proportional to the similarity between the nodes whereas in our model influence occurs only when similarity exceeds a hard threshold. Secondly, if influence does not occur, then a rewiring step necessarily does. This hard threshold also distinguishes our model from that of \cite{Vazquez2007}, although as shown below the qualitative behavior of both models is similar. It is worth pointing out that the rewiring step is designed such that the total number of links (or average degree) in the network is conserved. Also, the update rule as defined above always assumes that the node chosen first is the {\it influencer} and the node chosen second is the {\it adopter}. While most of the results described in this paper are restricted to this order in choosing the influencer and adopter, we discuss alternate orders of selection (motivated by the results in \cite{Nardini2008}) in Section~\ref{sec_order}.

\begin{figure}[!htbp]
\centerline{
\xsize
\epsfclipon
\epsfbox{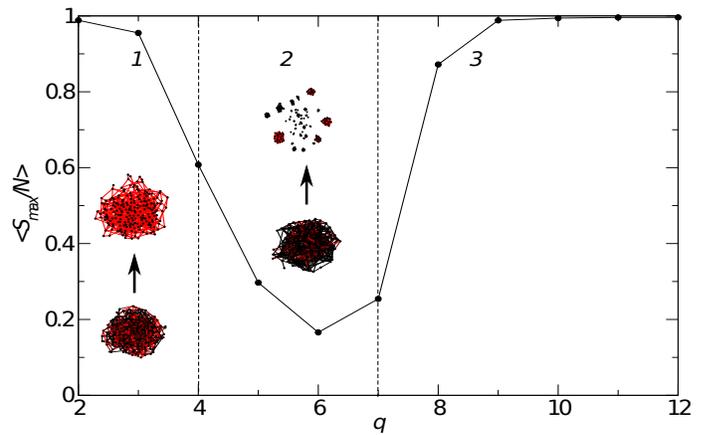}
}
\caption{The average size of the largest connected component (cluster),  $\langle S_{max}\rangle$, in the final state of the system as a function of $q$, starting from an ER Network with $N=200$, $\langle k\rangle=6.0$ and threshold $\phi=3$. The average is taken over $100$ realizations of network evolution. The plot shows three different phases characterized by distinct steady states of network structure (see text). This phase diagram is analogous to the one shown in Vazquez et al. \cite{Vazquez2007} for a related model. Also shown for phases $1$ and $2$ are initial (lower) and steady-state (upper) network snapshots for a single realization of co-evolution. The colors on the nodes represent the traits for a given attribute that is being tracked in the visualization. Snapshots in phase $1$ ($2$) have $q = 2$ ($q=7$) traits per attribute, and initially begin with each node having equal probability, $1/q$, of possessing any trait per attribute. We represent trait $0$ for the tracked attribute by red, and all other traits by black. Therefore, in the steady-state snapshot of phase $2$, each black cluster represents that all of its constituent nodes have adopted the same trait for the tracked attribute, but this common trait is different from $0$. }
\label{Fig1}
\end{figure}

\begin{figure}[!htbp]
\centerline{
\xsize
\epsfclipon
\epsfbox{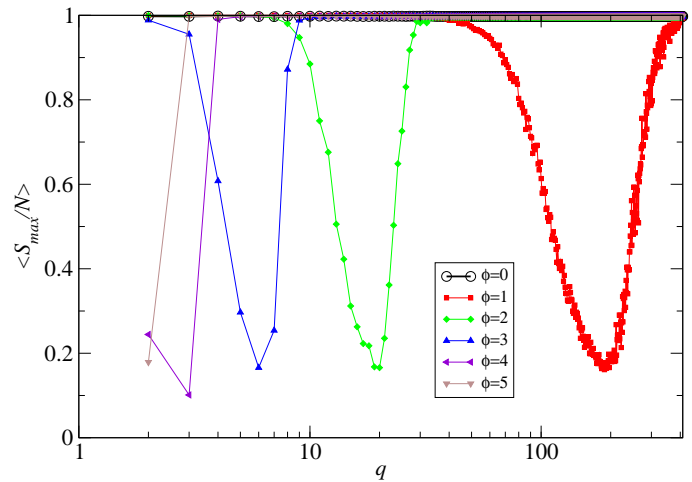}
}
\caption{Phase diagrams of the system as in Fig.~\ref{Fig1} for all possible values of $\phi$ (for $F=5$). As shown, the fragmentation transition takes place at successively higher values of $q$ as $\phi$ is decreased, $\to \infty$ as $\phi \to 0$. When $\phi$ is large (4 or 5) network is fragmented even for the smallest non-trivial value of $q(=2)$}
\label{Fig2}
\end{figure}

\begin{figure}[!htbp]
\centerline{
\xsize
\epsfclipon
\epsfbox{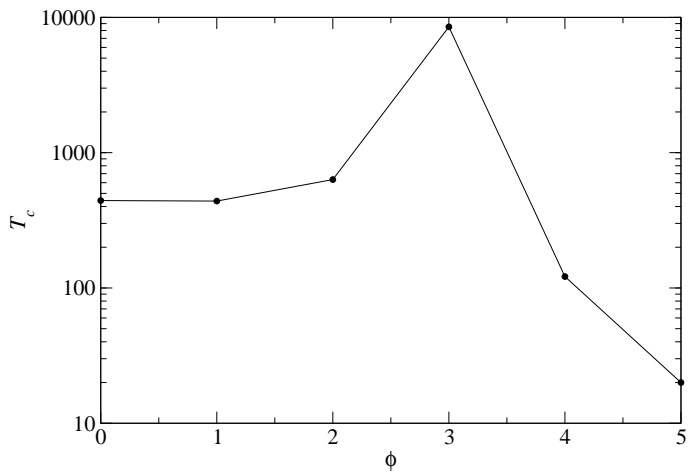}
}
\caption{Consensus time $T_c$ vs similarity threshold $\phi$ when the initial network is ER with $N=200$ and the average degree $\langle k\rangle =6.0$. $T_c$ varies non-monotonically with $\phi$ and reaches its maximal value at $\phi=3$.}
\label{Fig3}
\end{figure}

We first examine the effect of the number of traits per attribute $q$ in determining the steady state structure of the network, and confirm that our results are similar to those found in\cite{Vazquez2007,Klemm2003}. We fix $F = 5$ and set the similarity threshold to be $\phi = 3$. In particular, three phases differing in the steady state network structure are observed, as $q$ is 
varied, as shown in Fig.~\ref{Fig1}. In the first phase, observed for low values of $q$, the system evolves to a state where the network structure is static and global consensus is achieved i.e. for each attribute, each node possesses the same trait as all other nodes. As the number of traits is increased, at a critical value $q = q_c$, the system undergoes a phase transition to phase 2 where the steady state of the network consists of disconnected fragments with each fragment coming to  consensus locally. While the first two phases differ in the eventual network structure, they are similar in that the system eventually reaches a {\it frozen} state in which neither the node traits nor the network structure are evolving. However, further increasing $q$ beyond phase 2, eventually reveals a third phase where the initial dissimilarity among the nodes is so large that the system does not end up in a frozen state and rewiring continues indefinitely (Fig.~\ref{Fig1}). In this phase, at any given time, there exists a giant component; however, in the asymptotic limit of network size, the system will never reach consensus, so long as the average degree is not too small~\cite{Vazquez2007}. This three-phase behavior is expected to be seen for different choices of $F$ and $\phi$. For fixed $F$, as $\phi$ is increased the transitions occur at smaller values of $q$. For fixed $\phi$, as $F$ is increased the transition points move to higher values of $q$ (Fig.~\ref{Fig2}).
In order to study how the approach to consensus can be sped up, we continue with $F = 5$ and $\phi = 3$ and keep the number of traits per attribute $q$ fixed at $2$, so that the system is confined to phase 1, and a giant component reaching global consensus is guaranteed. Furthermore, the choice $\phi = 3$ results in the longest consensus times within phase 1, and thus represents the most challenging case within this phase, in the context of our study (Fig.~\ref{Fig3}).

\begin{figure}[!htbp]
\centerline{
\xsize
\epsfclipon
\epsfbox{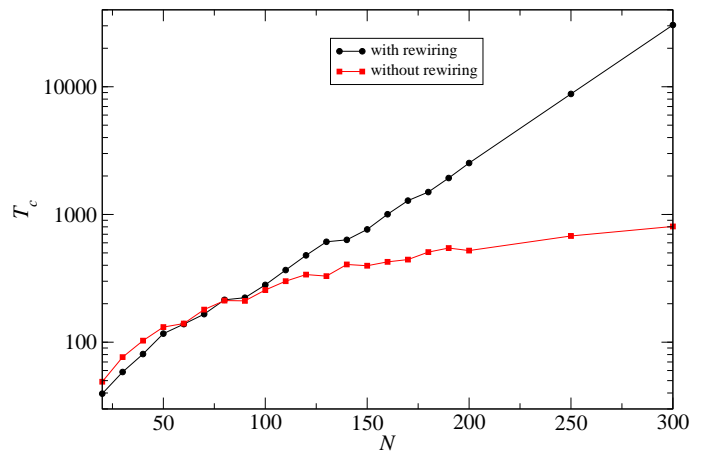}
}
\caption{Comparison between consensus times for two different cases $-$ when rewiring is present in the network (black circles) and when the network is static (red squares). These are results obtained from simulations, by averaging over $100$ realizations of network evolution. For each realization, the starting network is ER with  $\langle k\rangle=6.0$. Clearly, with rewiring $T_c$ scales exponentially with $N$ ( $T_c \sim \exp(\alpha N)$ where $\alpha \approx 0.02$) while without rewiring the scaling is linear.
}
\label{Fig4}
\end{figure}

For these parameters, when a pair of neighboring nodes is selected at time $t=0$, the probability of them meeting the similarity threshold, and hence the probability of influence, as one can see by elementary combinatorics, is exactly $\frac{1}{2}$ and equal to the probability of rewiring at time $t=0$. Since for the chosen parameters the system is in phase 1, the steady state reached is one where the network consists of a single connected component and the states of all nodes in this component are identical i.e. nodes are similar in the traits they possess for all $F$ attributes. Thus the steady state is one where a consensus is reached. Shown in Fig.~\ref{Fig4} is the scaling of the time to reach consensus (or consensus time) $T_c$ as a function of the system size. We contrast the behavior of a system where node attributes and network structure co-evolve to the behavior of a system where the node attributes evolve on a purely static network. The latter can also be thought of as the case when $\phi = 0$. As seen clearly, the effect of rewiring is detrimental to consensus times. With rewiring present, $T_c$ is exponential in $N$, in contrast to a linear scaling found for the static network. The divergence of $T_c$ with $N$ when rewiring is introduced can be qualitatively explained as follows. When a random node is chosen as the new end point of a rewired link, it is most probable that the chosen node has an attribute vector that is currently the most abundant vector - the majority state - in the population. If we assume that the attribute vector corresponding to the majority state does not change too frequently, then nodes in the majority state end up garnering a large number of rewired links. However, this comes at the detriment of the majority state, since when an influence step occurs and a random neighbor is chosen as the adopter, this neighbor is more likely to be in the majority state (due to the larger number of links leading to a node in the majority state). As a result, as soon as nodes in the majority state accrue more links than the rest of the population, they also become more likely to get influenced. This effect suppresses the growth of the majority state. Thus, the negative feedback due to rewiring strongly slows the spread of any particular state in the network.

\subsection{Committed Agents}
So far we have discussed how a consensus can be reached across all attributes i.e. the state vector of every node becomes identical. However, in a realistic scenario, it might be desirable to cause an entire population to adopt a given trait (opinion) for a given attribute (issue). For example, within a social network of teenage students in a high school, it is desirable that a consensus is reached where everyone views smoking as unhealthy behavior. Here, we study whether the introduction of committed agents \cite{Lu2009, Xie2011} can cause fast consensus on any given attribute characterizing the individuals. Without loss of generality we choose attribute $1$ as the one that committed agents intend to engineer a consensus on. We refer to this attribute as the {\it designated} attribute. Also, we assume that committed agents rigidly adopt trait $1$ for the designated attribute. Committed agents are thus considered un-influencable on attribute $1$, but are identical in their behavior to uncommitted nodes for other attributes. In the following, we redefine consensus time $T_c$ to mean the time taken for all nodes to adopt the trait proselytized by committed agents (i.e. $1$) for the first attribute. Attributes besides the designated one for committed nodes, as well as all attributes for uncommitted nodes are initialized to $1$ or $0$ with equal probability.

\begin{figure}[!htbp]
\centerline{
\xsize
\epsfclipon
\epsfbox{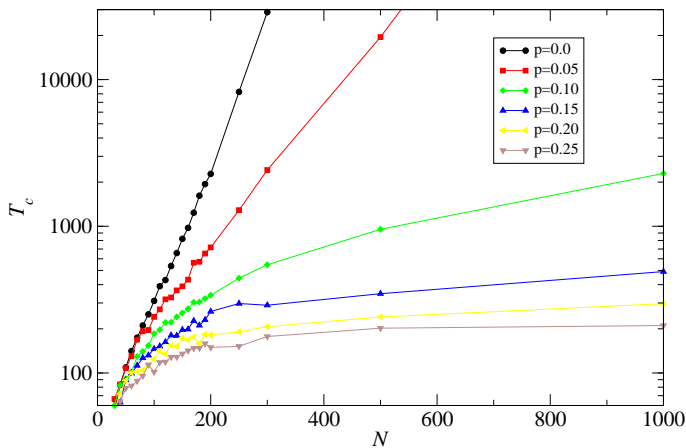}
}
\caption{Simulation results for consensus time $T_c$ as a function of network size $N$ when the initial network is ER with average degree $\langle k\rangle=6.0$ for different values of the committed fraction$p$. In these simulations, the influencer is chosen first, followed by the adopter. For low $p$, $T_c$ diverges exponentially with $N$. At the critical value $p_c \approx 0.1$, the system undergoes a transition beyond which $T_c \sim \log(N)$}
\label{Fig5}
\end{figure}

In Fig. ~\ref{Fig5}, we show the effect of introducing committed agents into the network. In particular, we choose randomly, a fraction $p$ of nodes as committed agents and study how the consensus time scaling with $N$ changes as $p$ is varied.  As expected, for $p=0$ the exponential scaling of $T_c$ with $N$ is recovered. Although consensus time values decrease for any given $N$ as $p$ is increased, the scaling behavior remains unchanged until a critical fraction of $p = p_c \approx 0.1$ is reached. Beyond this critical fraction, consensus times dramatically change their scaling behavior and become logarithmic in $N$. As mentioned in the introduction, a similar result was found in \cite{Xie2011, Zhang2011} for a different model of social influence on a structurally static network. This result indicates that beyond a critical value of the committed fraction, this fraction can efficiently overcome the resistance to consensus generated by the random rewiring taking place in the network. Fig. ~\ref{Fig6}(a) shows consensus time as a function of $p$, for different network sizes.

\section{finite size scaling analysis}
Here we employ finite-size scaling analysis to obtain an estimate of the critical committed fraction $p_c$ from our simulation results. In parlance with the theory of phase transitions, we assume the following scaling ansatz for $T_c$:
\[ T_c(p,N) = N^\alpha f(N^\beta (p-p_c))\]
where $p_c$ is the critical value of $p$ and $\alpha$ and $\beta$ are critical exponents. The function $f$ is an unknown scaling function.
\begin{figure}[htbp]
\centerline{
\xsize
\epsfclipon
\epsfbox{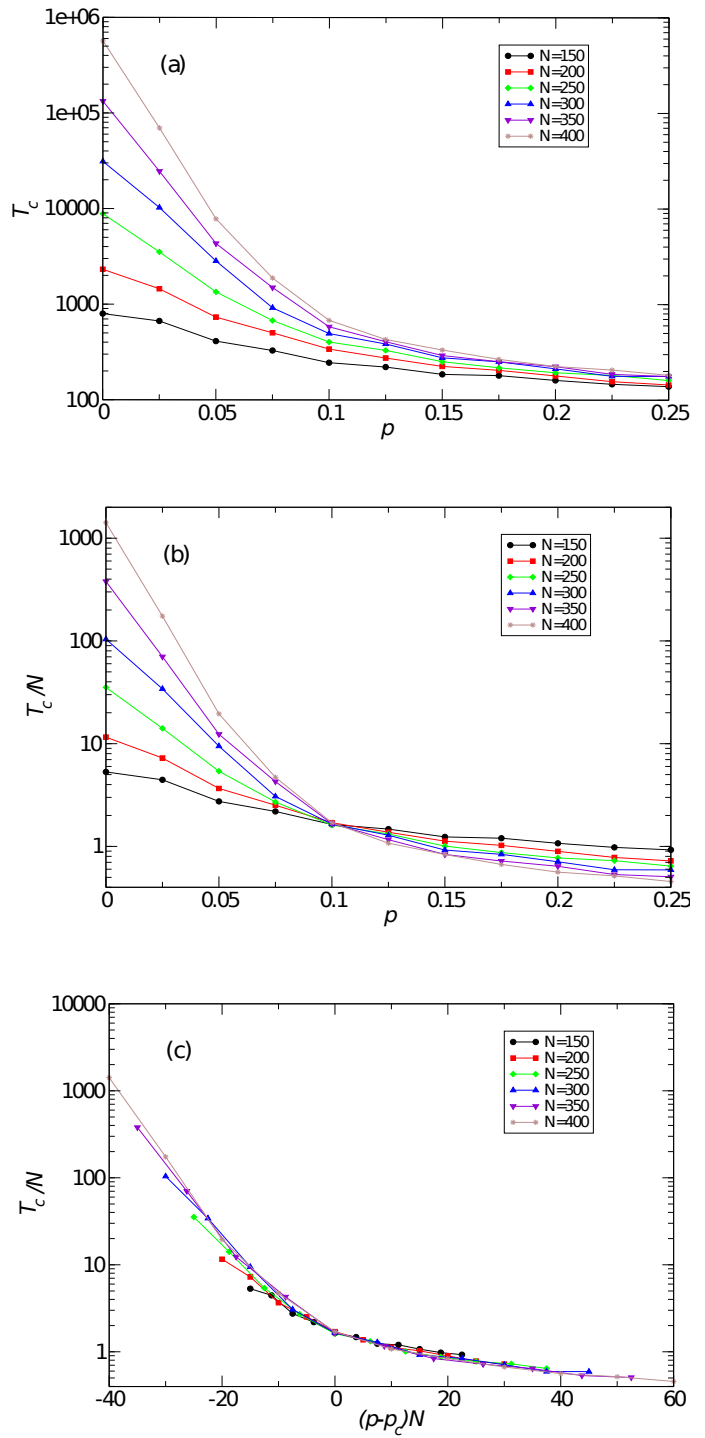}
}
\caption{(a) $T_c$ vs $p$ starting from an ER network with average degree $\langle k\rangle=6.0$ for different system sizes $N$. (b)$\frac{T_c}{N^\alpha}$ with $\alpha=1.0$ vs $p$ for the data in (a). The critical point - the value of $p$ at which the curves intersect - is $p_c\approx0.1$ (see text). (c) $\frac{T_c}{N^\alpha}$ ($\alpha=1.0$) vs $N^\beta(p-p_c)$ shows good scaling collapse in the vicinity of $p_c$ for $\beta=1.0$.}
\label{Fig6}
\end{figure}

At $p = p_c$, the scaling ansatz yields $T_c N^{-\alpha} = f(0)$, which implies that if $T_c N^{-\alpha}$ is plotted as a function of $p$, curves for different $N$ values must intersect at $p = p_c$. Since $\alpha$ itself is unknown, we progressively increase $\alpha$ from values close to zero until curves of $T_c N^{-\alpha}$ versus $p$ corresponding to different $N$, intersect at a single point. We find that a single intersection point is obtained for $\alpha = 1$ and this point is at $p = p_c \approx 0.1$ (see  Fig.~\ref{Fig6})(b). To obtain the exponent $\beta$, we plot $\frac{T_c}{N^\alpha}$ as a function of $N^\beta(p-p_c)$, and vary $\beta$ until a good scaling collapse is found for all curves near $p_c$. Following this procedure we find $\beta \approx 1$. The scaling collapse is shown in Fig.~\ref{Fig6}(c).

\section{Effect of influencer-adopter selection order} \label{sec_order}

In this section we study how the order of selecting the influencer and adopter nodes affects consensus times, and also the transition observed in consensus times as the committed fraction is varied. Surprisingly, such subtle changes in selection order can have a profound effect on consensus times.
\subsection{Adopter-first selection}

\begin{figure}[!htbp]
\centerline{
\xsize
\epsfclipon
\epsfbox{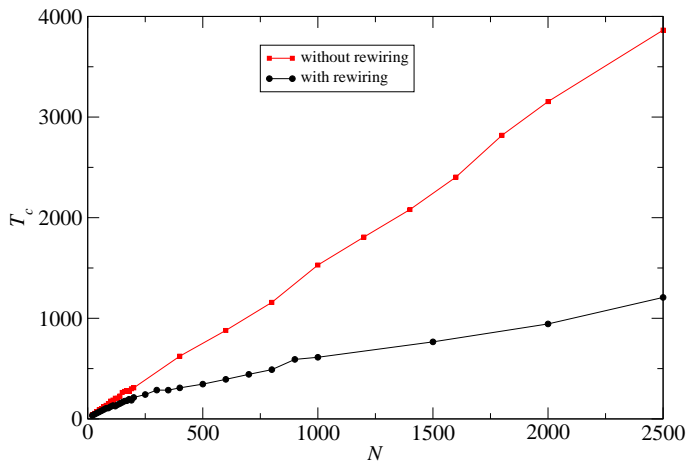}
}
\caption{The consensus times for two different cases $-$ when rewiring is allowed in the network (black circles) and when the network is static (red squares) when first chosen node is the adopter (and $p=0.0$). The initial network is ER with $\langle k\rangle=6.0$. $T_c$ on a  static network scales linearly with $N$ but with rewiring $T_c$ follows a power law with exponent $\approx 0.7$.}
\label{Fig7}
\end{figure}

\begin{figure}[htbp]
\centerline{
\xsize
\epsfclipon
\epsfbox{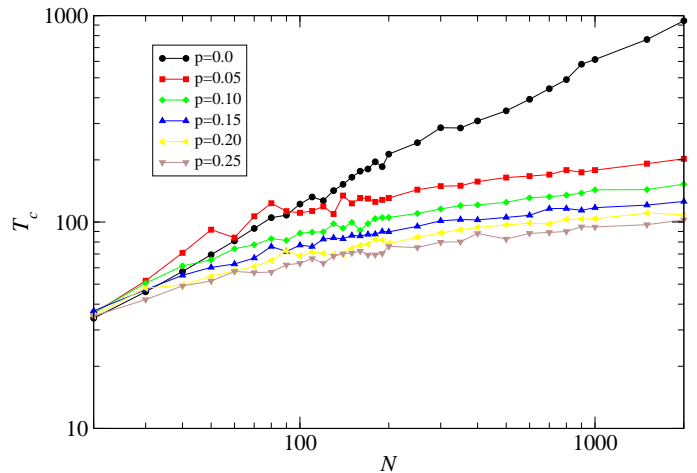}
}
\caption{Scaling of consensus time, $T_c$, as a function of network size, $N$, for different values of the  committed fraction $p$ for adopter-first dynamics. In the absence of committed agents $T_c \sim N^{0.7}$ and is logarithmic in $N$ for any $p>0$.}
\label{Fig8}
\end{figure}

In adopter-first selection, a random node $i$ is selected, followed by a randomly chosen neighbor $j$ of $i$. However in contrast to the model studied so far, if the criterion for an influence step to occur is met, then $i$ adopts the trait of $j$ for a randomly chosen attribute from among those for which they currently do not share the same trait. Thus, as the name suggests, the first chosen node is treated as the adopter and the second as the influencer. A comparison between consensus times on static networks and networks with rewiring under this updating scheme (Fig.~\ref{Fig7}) suggests that rewiring generates a positive feedback which aids the network in reaching consensus. This is in contrast to what was observed in the previous case of influencer-first selection.
Surprisingly, for adopter-first dynamics, the previously observed exponential divergence of  $T_c$  with $N$ is absent even in the case when the committed fraction $p = 0$. Instead, with $p = 0$, consensus time grows as a power law in $N$, with an exponent $\approx 0.7$. Furthermore, for $p > 0$, $T_c \sim \log N$ suggesting the existence of a transition in the scaling behavior of $T_c$ occurring very close to $p= 0$ (Fig.~\ref{Fig8}).

\subsection{Unbiased selection}
\begin{figure}[!htbp]
\centerline{
\xsize
\epsfclipon
\epsfbox{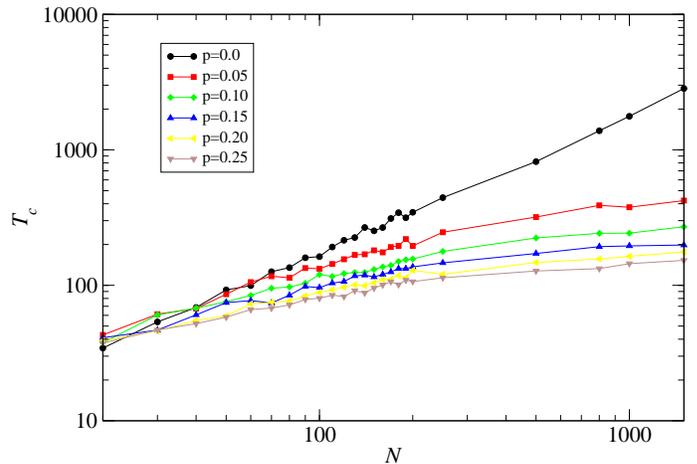}
}
\caption{Scaling of consensus time $T_c$ with network size $N$ for different values of committed fraction $p$ for unbiased dynamics. For these simulations, a microscopic update consisted of randomly picking a link, and then randomly selecting one of the endpoint nodes of the chosen link to be the adopter and the other, the influencer. For $p=0$, $T_c$ follows a power law with an exponent $\approx1$ (linear) and for any $p>0$, $T_c \sim \log(N)$.}
\label{Fig9}
\end{figure}

In the selection orders considered so far, the direction of influence is always fixed beforehand. Here, we investigate the case in which, after a pair of neighboring nodes is selected, a random one among them is chosen to be the influencer and the other, the adopter (provided the criterion for an influence step is met). In other words, a pair of nodes $(i,j)$ is selected, and with probability $\frac{1}{2}$ node $i$ adopts one of node $j$'s traits, and otherwise node $j$ adopts one of node $i$'s traits. In this case also, a transition in scaling behavior of $T_c$ appears when $p$ is  changed from zero to a non-zero value. However, the transition here is from a linear scaling, $T_c \sim N$ for $p = 0$, to $T_c \sim \log N$ for $p > 0$ (Fig.~\ref{Fig9}).

\section{Conclusions}
Using a variant of the Axelrod model with homophilic rewiring, we have studied how a small fraction of committed agents can dramatically influence the scaling of consensus times on structurally evolving networks. So far, the vast majority of models studying how a targeted change in opinion or behavior can be engineered have been confined to networks that are fixed in their topology. By considering  the effect of persistent opinion holders - committed agents - on structurally evolving networks, we show that introducing a committed fraction $p > p_c$ represents a scalable method to cause widespread adoption of a given opinion on such networks. We have also considered variations to the update rule involving the selection order of influencers and adopters and shown that the transition in scaling behavior of $T_c$ across some $p_c$ is a consistent feature across these variations, even though the precise scaling behavior is dependent on the selection order. Moreover, our results show that in the worst case scenario for consensus time - the influencer-first case - the introduction of a critical number of committed nodes can result in a dramatic reduction in consensus time. While we have employed a network model with simple but plausible rewiring and influence rules, with data on time-evolving networks becoming increasingly available, it may be worthwhile studying empirically how the structural evolution of such networks is coupled to the attributes associated with their constituent nodes.

\section{acknowledgments}
This work was supported in part by the Army Research Laboratory under Cooperative Agreement Number W911NF-09-2-0053, and by the Office of Naval Research Grant No. N00014-09-1-0607. The views and conclusions contained in this document are those of the authors and should not be interpreted as representing the official policies, either expressed or implied, of the Army Research Laboratory or the U.S. Government.  The U.S. Government is authorized to reproduce and distribute reprints for Government purposes notwithstanding any copyright notation here on. We also thank C. Lim, J. Xie, A. Asztalos, A. Goscinski and D. Hunt for useful comments and discussions.


\end{document}